%% file: nanowires.tex
\documentclass[%
 reprint,
 amsmath,amssymb,
 aps,
 pre,
floatfix,
]{revtex4-2}

\usepackage{graphicx}% Include figure files
\usepackage{dcolumn}% Align table columns on decimal point
\usepackage{bm}% bold math
\usepackage{csquotes}

\begin{document}

\preprint{APS/123-QED}

\title{Analytical modeling of orientation% 
\\ effects in random nanowire networks}% Force line breaks with \\

\author{Milind Jagota}
 \email{Corresponding author. Contact at mjagota@stanford.edu}
\author{Isaac Scheinfeld}%
 \email{ischeinfeld@stanford.edu}
\affiliation{%
 Stanford University \\
 Stanford, CA 94305
 }%

\date{\today}% It is always \today, today,

\begin{abstract}
Films made from random nanowire arrays are an attractive choice
for electronics requiring flexible transparent conductive films. 
However, thus far there has been no unified theory for predicting
their electrical conductivity.
In particular, the effects of orientation distribution on
network conductivity remain poorly understood.
We present a simplified analytical model for random 
nanowire network electrical conductivity that is 
the first to accurately capture the effects of 
arbitrary nanowire orientation distributions on conductivity.
Our model is an upper bound and converges to the true
conductivity as nanowire density grows. 
The model replaces Monte Carlo sampling with an asymptotically
faster computation and in practice can be computed much more 
quickly than standard computational models. 
The success of our approximation provides novel 
theoretical insight into how nanowire orientation affects
electrical conductivity, illuminating directions for future
research. 
\end{abstract}

\maketitle

\section{\label{sec:level1}Introduction}

Transparent conductive films are a crucial component of touch
screens and solar cells, among various other electronics
\cite{kumar2010race, gordon2000criteria}.
One approach to making transparent conductive films that has
been widely studied and deployed is to randomly disperse highly
conductive nanowires into a substrate. 
Films made in this way, using conductive material such as silver
nanowires or carbon nanotubes, display competitive electrical and
optical properties to alternatives, while being cheaper and more
flexible than the performance standard Indium Tin Oxide
\cite{choi2019high,hecht2011emerging,hu2011metal,%
      lee2008solution,teymouri2017low}. 
The latter property is particularly valuable as flexible electronics
continue to become more mainstream in consumer devices. 
However, despite the wide interest in applying them, there is no
unified theory for predicting electrical properties of random
nanowire networks, and many observed effects have not been
fully characterized or explained. 
As a result, the technology remains underdeveloped, and there
is undoubtedly still room for improvements in performance. 

The majority of results describing properties of
random nanowire networks have been experimental or via
direct computational simulation.
Various studies have experimentally compared electrical
properties of films using different conductive rods, such as
silver nanowires and carbon nanotubes \cite{hecht2011emerging,%
hu2011metal, marus2015comparative}. 
Agreement between simulation and experimental observations of
electrical properties has also been well established for the
classes of random nanowire networks that are easiest to produce
experimentally \cite{mutiso2013integrating}. 
More recently, computational models have been used to maximize
electrical performance of random nanowire networks by varying
the distributions from which the networks are sampled
\cite{jagota2015conductivity, behnam2007effects,%
      behnam2007computational, du2005effect,%
      white2009simulations, tarasevich2018simulation,%
      pimparkar2007limits, hicks2018effect}. 
Some of these results have been verified experimentally
\cite{bellew2015resistance, ackermann2016effect,%
      marus2017effect, wu2016aligned}. 
In particular, various computational studies have
demonstrated that it is possible to improve electrical
conductivity of nanowire networks by controlling nanowire
orientation \cite{jagota2015conductivity, 
behnam2007effects,behnam2007computational, du2005effect,%
white2009simulations, tarasevich2018simulation,% 
pimparkar2007limits}. 
However, this effect is not well understood
and there is no simple framework to predict the result of using 
a specific, arbitrary orientation distribution.

Recently, a number of analytical models have also been 
developed to describe properties of random nanowire 
networks, but none thus far have explained the effect 
of nanowire orientation on electrical conductivity 
in full generality  
\cite{marus2017towards, kim2018systematic, % 
tarasevich2019electrical, forro2018predictive, % 
manning2019electro, ponzoni2019contributions, % 
benda2019effective, kumar2017current}.
Forro \textit{et al}. proposed a model derived assuming
high nanowire density, so that potential drop across
nanowire networks can be assumed to be linear 
\cite{forro2018predictive}. The model is accurate in the
high density regime and yields a closed-form expression. 
Benda \textit{et al}. obtained a closed form expression 
for network conductivity by numerically fitting a 
physically interpretable form to Monte Carlo simulations,
while Manning \textit{et al}. developed a theoretical 
framework for analyzing both electrical and optical 
performance of nanowire networks 
\cite{benda2019effective, manning2019electro}. 
However, these models are developed under the
assumption of uniformly distributed wire orientation and do not
generalize in a clear manner to random orientation of an arbitrary
distribution. 

In this work, we present the first analytical model for random
nanowire network conductivity that accurately captures the 
effects of arbitrary distributions of nanowire orientation.
Our approximate model replaces Monte Carlo sampling with an
asymptotically less expensive computation and is empirically much 
faster than standard computational models.
It approaches the limiting dependency of network conductivity
on nanowire density, with small errors even at moderate nanowire 
densities.
Furthermore, the structure of our approximations provides
novel intuition for how orientation affects network conductivity
as well as intuition for the behavior of random nanowire networks
in general.

\section{Model Construction}

\subsection{Setting}
We begin by presenting the setting in which we develop our model.
We consider networks comprised of 1-dimensional nanowires (linear, 
widthless sticks) inside a square space of unit length in each 
direction with periodic boundary conditions at the top and bottom.
To simplify notation, we assume nanowires have fixed length $l$,
but our approach generalizes naturally to having a random
distribution over wire length.
Each nanowire is described by an $(x,y)$ coordinate pair and an
angle $\theta$, where the coordinate pair represents the location
of the wire center and $\theta$ is the angle relative to the horizontal.
The coordinates and the angle are sampled randomly, where all
values are assumed independent and each nanowire in a network
is assumed to be independent. 
We denote the sampling distributions of $x,y,\theta$ by
$\mathcal{X}, \mathcal{Y}, \Theta$, respectively.

The primary electrical property of interest for random nanowire
networks is the sheet conductivity $\sigma$, which is a random variable. 
Sheet conductivity transitions sharply from being zero with
overwhelming probability to being greater than zero with overwhelming
probability at a particular number of nanowires that is a function of 
$l$, known as the percolation threshold 
\cite{stauffer2018introduction}. 
The dimensionless quantity 
\begin{align}
    C_N := N|l|^{2}
    \label{normalized_conctration}
\end{align}
where $N$ is
the number of nanowires in an network and $|l|$ indicates the 
wire length normalized by dividing by box width, is often used 
as a normalized concentration of nanowires because it allows 
direct comparison to the percolation threshold
\cite{jagota2015conductivity}.
We assume that our nanowire networks are well above the
percolation threshold so that they are guaranteed to have conductivity
greater than zero.
We focus on modeling the expected value of the sheet conductivity   
$\mathrm{E}\sigma$, because the variance of sheet conductivity is typically 
small relative to the expected sheet conductivity for large $N$
\cite{mutiso2013integrating, jagota2015conductivity}.

Figure 1 displays how the sheet conductivity is physically
defined, using a network sampled with nanowire positions and orientations
both distributed uniformly.
We place electrodes at the left and right boundary of the
network ($x=0$ and $x=1$) and calculate the current when
1 volt is applied.
This current can then be used to calculate the sheet conductivity. 
In general, there are three sources of resistance in nanowire
networks which determine the conductivity along with the geometry.
These three sources are the resistance of wires themselves,
the resistance at the junctions between two wires, and the
resistance at the junctions between a wire and an electrode.
In many real nanowire networks, the wire resistance is small
compared to the resistance at junctions \cite{mutiso2013integrating}.
We assume that this is the case and choose to ignore the
wire resistance moving forward.
However, our method can be generalized to account for wire
resistance, and we discuss this in Section V.
We set the resistance between two wires to be a constant
$1\Omega$ and set the resistance between a wire and an
electrode to be a constant $\frac{1}{100}\Omega$.
The conductivity of a particular network is determined solely by the
ratio between these two quantities up to scaling.
We expect the wire-wire resistance to be multiple orders
of magnitude larger than the wire-electrode resistance, and
these quantities are thus reasonable.

\begin{figure} 
\includegraphics[width=0.8\linewidth]{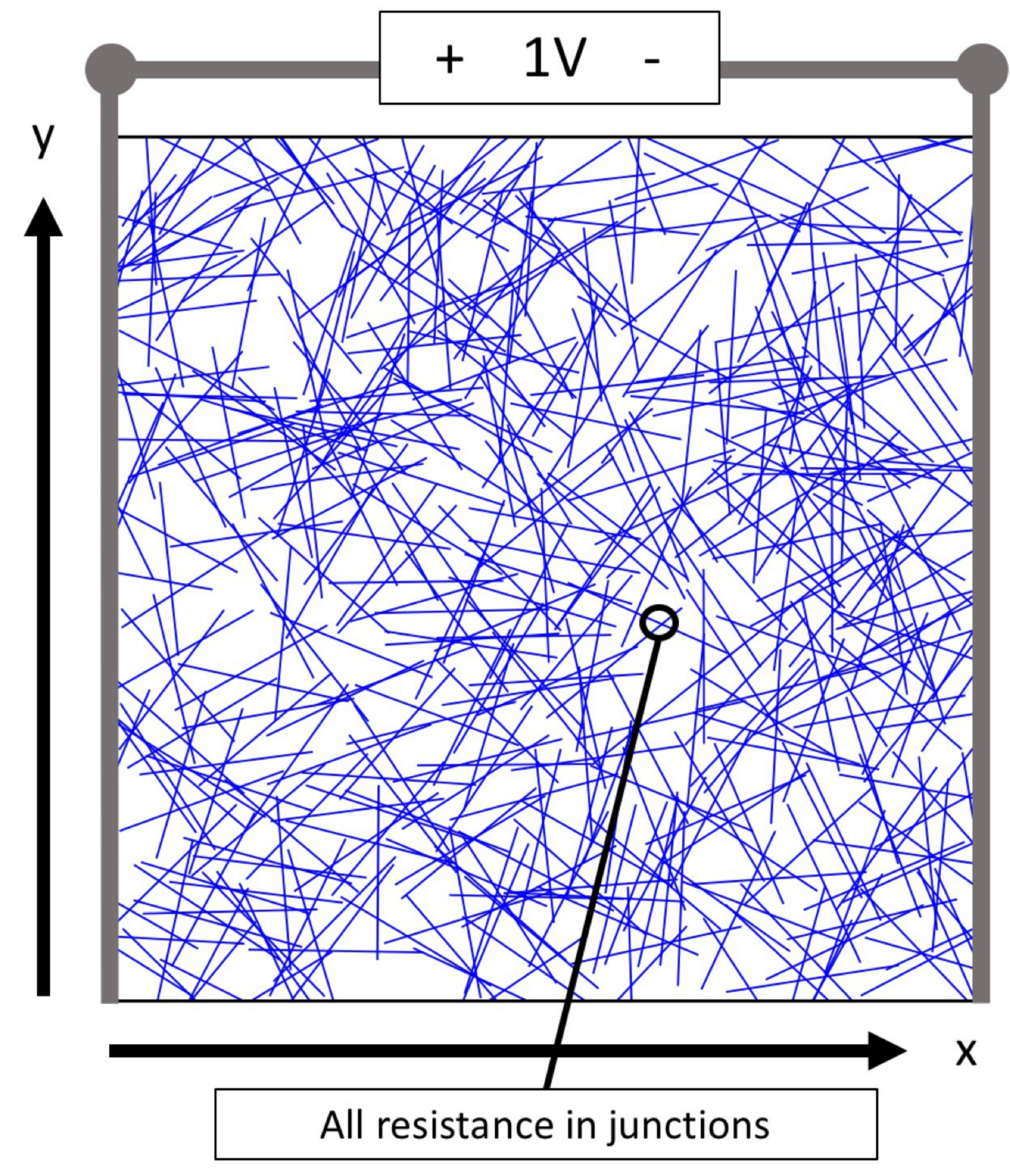}
\caption{The sheet conductivity of a nanowire network is calculated
by computing the current when 1 Volt is applied by
electrodes spanning the left and right border of the network.
We assume that nanowire network resistance is dominated by
junction resistance and ignore wire resistance. The $x$ direction
is defined as the direction of current flow, and the $y$ direction
is perpendicular.} 
\end{figure}

In this setting, $\sigma$ can be calculated exactly for a
particular network from the symmetric $(N+2) \times (N+2)$
adjacency matrix of the electrical network, which we denote as $A$. 
The first $N$ rows of this matrix each correspond to a
single wire, while the last two rows correspond to the
left and right border electrode.
An off-diagonal element of the matrix is 1 if the two
corresponding objects touch, and all diagonal elements are 0.
From $A$, we can use the two resistance values to construct the
Laplacian matrix of the nanowire network $L$, of the same shape as $A$.
This is the matrix that, when multiplied by the vector of
node voltages $V$, gives the vector of node net current flow 
$J$ as given in Eq.~(\ref{kirchoff_equation}), and is a 
linear function of $A$ \cite{klein1993resistance}.
\begin{equation}
   LV = J
    \label{kirchoff_equation}
\end{equation}
We can then calculate the current flowing from the left 
electrode by setting the voltages at the left and right electrodes 
in the vector $V$ and solving for the remaining voltages.
Dividing this current by the applied voltage yields the sheet
conductivity \cite{klein1993resistance}.

\subsection{Model definition}
The expected sheet conductivity $\mathrm{E}\sigma$ has
most often been studied by direct sampling of nanowire networks
\cite{mutiso2013integrating, jagota2015conductivity,%
      behnam2007effects,%
      behnam2007computational, du2005effect,%
      white2009simulations, tarasevich2018simulation,%
      hicks2018effect, pimparkar2007limits}. 
This procedure involves numerous steps.
For each network, $N$ nanowires are sampled according
to the distributions $\mathcal{X}, \mathcal{Y}, \Theta$.
Then, the adjacency matrices $A$ for the networks are generated.
From these matrices, observations of the sheet conductivity
can be calculated by applying Kirchhoff's Laws,
which are then averaged to yield an estimate.
We denote this empirical estimate by $\hat{\sigma}$, defined in 
Eq.~(\ref{empir_est}), where $\sigma(A_i)$ refers to the sheet 
conductivity of the network represented by the adjacency matrix $A_i$.
\begin{equation}
    \hat{\sigma} = \frac{1}{M}\sum_{i=1}^{M}{\sigma(A_i)}
    \label{empir_est}
\end{equation}

While this approach converges rapidly to $\mathrm{E}\sigma$ as the
number of sampled networks $M$ increases, it has a number of drawbacks.
First, it is slow: calculating the adjacency matrix $A$ from a
list of wire coordinates and angles requires checking all pairs
of nanowires for intersection, as well as computing a Cholesky
decomposition of an $N \times N$ matrix. 
While there are methods to speed up both of these steps, the
procedure is still at least $O(MN^2)$ and so collecting many
samples for high conductivity films is slow.
In addition, this sampling based procedure makes 
interpretation of observed effects difficult, which
limits physical intuition. 

An exact analytical model for the sheet conductivity would
fix these issues, but directly deriving an expression for
$\mathrm{E}\sigma$ is very difficult even under the simplest
distributions $\mathcal{X}, \mathcal{Y}, \Theta$.
A common approximation for this type of problem is to move the
expectation inside of the complicated function, as shown in
Eq.~(\ref{naive_approach}).
The right side of this equation is defined by treating $\mathrm{E}A$
as a weighted adjacency matrix; the Laplacian $L$ is constructed 
from $\mathrm{E}A$ by the same linear relationship as for an 
ordinary adjacency matrix $A$, and the sheet conductivity is 
calculated by solving the same matrix equation involving $L$.
\begin{equation}
    \mathrm{E}\sigma(A) \approx \sigma(\mathrm{E}A)
    \label{naive_approach}
\end{equation}
However, this naive approach fails catastrophically for random
nanowire networks.
None of the spatial structure of the networks is captured because
all nanowires are indistinguishable according to $\mathrm{E}A$.
Using $\sigma(\mathrm{E}A)$ as a model results in a massive
overestimate of the sheet conductivity that is not useful.

To develop our analytical model, we modify the approach of
moving the expectation inside the function to directly capture
spatial structure of random nanowire networks.
We first observe that $\sigma$ is clearly invariant to reindexing
the wires in a network and recalculating the adjacency matrix
$A$ accordingly. 
We choose to assume, without loss of generality, that the wires
are always reindexed according to increasing $x$-coordinate.
Specifically, define the random matrix $A^*$ as 
\begin{equation}
    A^*_{\mathrm{rank}(i),\mathrm{rank}(j)} = A_{ij} 
    \label{A_star}
\end{equation}
where the function $\mathrm{rank}(i)$ gets the placement of
$x_i$ in the list of $x$-coordinates when sorted from smallest
to largest and leaves the electrode indices fixed.
Our approximate model $\sigma^*$ is then defined in 
Eq.~(\ref{approx_model}), where $\sigma(\mathrm{E}A^*)$ is 
defined in the same way as the right side of Eq.~(\ref{naive_approach}).
\begin{equation}
    \sigma^* := \sigma(\mathrm{E}A^*)
    \label{approx_model}
\end{equation}

Under slightly more restrictive assumptions, we can prove that
$\sigma^*$ is greater than $\mathrm{E}\sigma$ for all
$\mathcal{X}, \mathcal{Y}, \Theta$ using Jensen's inequality;
details are presented in Section II.C. 
Despite being an upper bound, $\sigma^*$ is able to capture
the dependency of conductivity on both wire concentration
and orientation distribution due
to the choice of assumed wire permutation; $\mathrm{E}A^*$
encodes most of the spatial structure of the networks.
We illustrate this property of our model in Figure 2 by plotting
the values of a random sorted adjacency matrix $A^*$ as well
as the values of $\mathrm{E}A^*$ under the same distributions.
Due to the sorted order that is assumed, the matrices 
$A^*$ for sampled random networks are banded, because wires
near in index are also near in $x$-coordinate and therefore more 
likely to intersect. 
The expected adjacency matrix $\mathrm{E}A^*$ reproduces this 
key property well. 

\begin{figure} 
\includegraphics[width=0.8\linewidth]{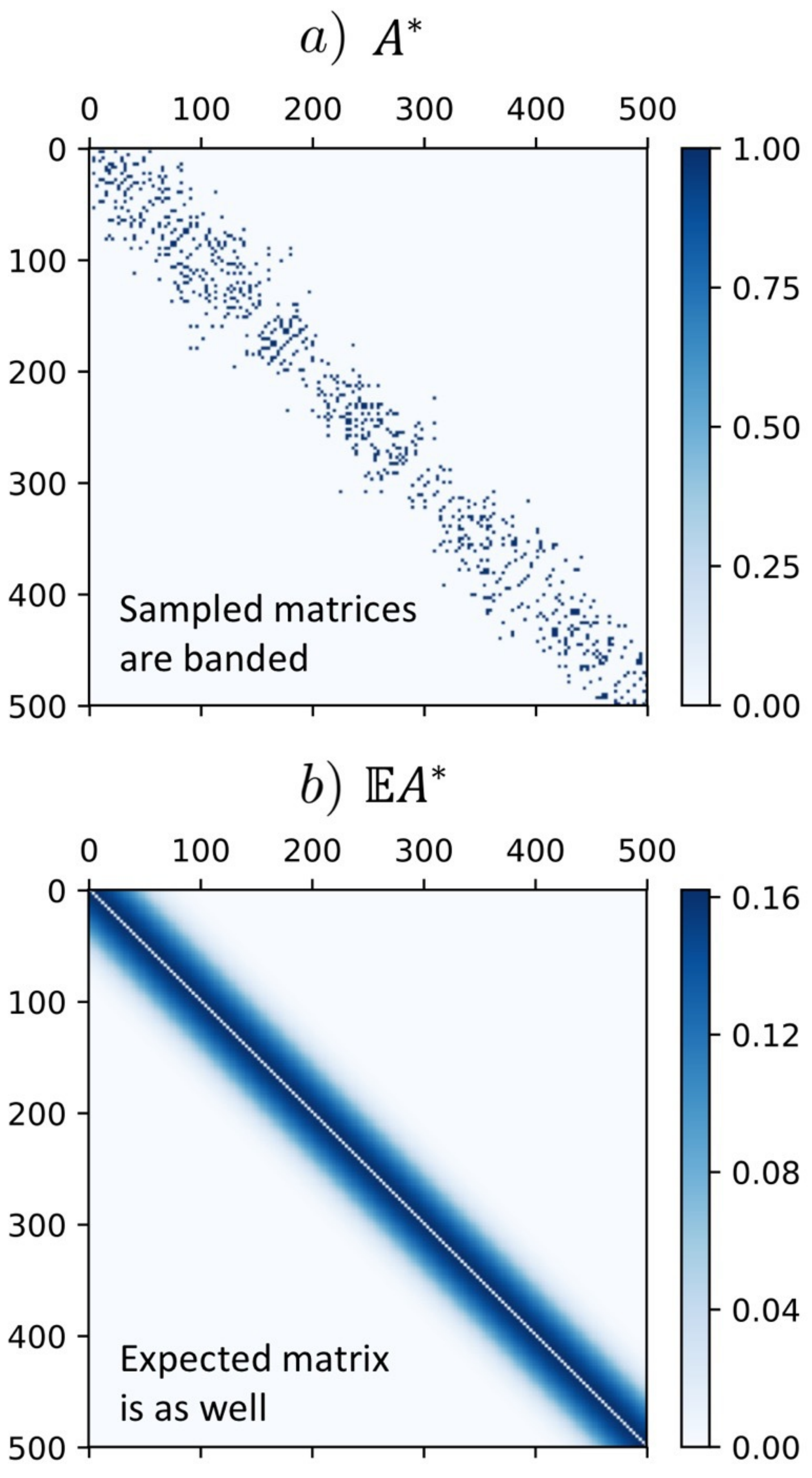}
\caption{We plot the values of (a) the sorted
adjacency matrix $A^*$ of a random nanowire network, and (b)
the expected adjacency matrix $\mathrm{E}A^*$ for the distributions
that the sample was drawn from, with $N=500$ and $l=0.2$. The
expected adjacency matrix captures the banded structure of the 
sampled matrices $A^*$. Note that $\mathrm{E}A^*$ has a small
maximum value because even if two nanowires have no $x$-separation,
the probability of them intersecting is small when $l$ is small.} 
\end{figure}

In the true system, nanowires that intersect are close in 
$y$-coordinate as well as in $x$-coordinate.
We encode this effect only with respect to $x$-coordinate and not
$y$-coordinate, but our empirical results verify that our model 
is useful regardless.
This result has interesting implications which we discuss
in Section IV.

\subsection{Proof that $\sigma^*$ is an upper bound}
We argue that $\sigma^*$ is an upper bound on $\mathrm{E}\sigma$
under slightly more restrictive assumptions. 
Note that the Laplacian matrix $L$ for a particular nanowire
network is a linear function of the sorted adjacency matrix 
$A^*$ \cite{klein1993resistance}. 
It satisfies the Kirchhoff current equation given in
Eq.~(\ref{kirchoff_equation}), where $V$ is the vector of voltages
at each of the $N+2$ objects and $J$, the net current flowing
into each node, is zero at all nodes other than the electrode nodes.
To reduce notation for units, we assume in this section that $V$ is 
made dimensionless by dividing each element by $1\mathrm{V}$. 
$L$ and $J$ then both have units of inverse resistance.

Under the assumed normalization of $V$, the sheet conductivity is 
equal to the current flowing out of the left border electrode 
(node $N+1$) when we set the voltage at the left border to be 
1V and the voltage at the right border to be 0V 
($V_{N+1}=1$, $V_{N+2}=0$).
With these values of $V$ set, the Kirchhoff current equation
is given by Eq.~(\ref{current_law_elim}).
\begin{equation}
   L_{1:N,1:N}V = -L_{1:N,N+1}
    \label{current_law_elim}
\end{equation}
We use the notation $B_{i:j,k:l}$ to refer to the
submatrix of $B$ from rows $i$ to $j$ and columns $k$ to $l$.
A single index indicates taking a single row or column.

We proceed by adding two minor assumptions. 
We first assume that for fixed 
distributions $\mathcal{X}, \mathcal{Y}, \Theta$, the number of 
nanowires crossing the left electrode is a constant integer $M$. 
For the high density networks we study, the variance of 
this quantity is small with respect to its expected value 
and does not cause much variance in sheet conductivity. 
Second, we assume that the $M$ nanowires that
cross the left border are the first $M$ indices in $A^*$. 
Under the sorting that is used for $A^*$, this is
the most likely set of $M$ wires to cross the left border, and the
variance of these indices also does not cause much variance in sheet 
conductivity. This assumption can be viewed
as a definition of sheet conductivity where
we attach our left electrode to the leftmost $M$ wires
based on center location, as opposed to based on left endpoint location.

Under these assumptions, the sheet conductivity is given by
Eq.~(\ref{alt_sheet_cond}), where $R_{ew}$ is the wire-electrode 
resistance in Ohms and $E_M$ is the $N$ 
dimensional vector that is 1 in the first $M$ elements and 0
otherwise.
\begin{eqnarray}
    \sigma &&= \frac{1}{R_{ew}}\sum_{i=1}^{M}%
    (1 - V_i) \nonumber \\
    && = \frac{1}{R_{ew}}(M - \sum_{i=1}^{M}%
    V_i) \nonumber \\
    && = \frac{1}{R_{ew}}(M - E_M^T%
    V) \nonumber \\
    && = \frac{1}{R_{ew}}(M + (E_M^T% 
    (L_{1:N,1:N})^{-1}L_{1:N,N+1})
    \label{alt_sheet_cond}
\end{eqnarray}
The inverse of $L_{1:N,1:N}$ exists when the network is
connected, which is true because we assume that our networks are 
well above the percolation threshold.

Since the first $M$ nanowires cross our left
measurement electrode, $L_{1:N,N+1}$ is given by
\begin{equation}
   L_{1:N,N+1} = -\frac{1}{R_{ew}}E_M
    \label{right_side}
\end{equation}
We can use this value to write another expression for
$\sigma$ in Eq.~(\ref{sheet_cond}).
\begin{equation}
   \sigma = \frac{1}{R_{ew}}(M - %
   \frac{1}{R_{ew}}E_M^T (L_{1:N,1:N})^{-1}E_M)
    \label{sheet_cond}
\end{equation}

Since $M$ is assumed to be constant, the only randomness in
$\sigma$ comes from $L_{1:N,1:N}$.
Since this matrix is positive definite when the network is connected
and is a linear function of $A^*$, $\sigma$
is a concave function of $A^*$. 
Jensen's inequality then tells us that for all
$\mathcal{X},\mathcal{Y},\Theta$, $\sigma^*$ is an upper
bound on $\mathrm{E}\sigma$ as shown in Eq.~(\ref{jensen}).
\begin{eqnarray}
    \mathrm{E}\sigma(A) = \mathrm{E}\sigma(A^*) %
    \leq \sigma(\mathrm{E}A^*) = \sigma^*
    \label{jensen}
\end{eqnarray}

\section{Model computation}
\subsection{Methods for computing $\mathrm{E}A^*$}
The approximate model $\sigma^*$ is useful because we can directly
compute $\mathrm{E}A^*$ in a wide variety of circumstances.
This eliminates the need for Monte Carlo sampling of networks and 
solving a linear system of equations for each sample.
Here we present a method for computing $\mathrm{E}A^*$ when 
$\mathcal{X}, \mathcal{Y}$ are the uniform distribution, which 
is an assumption used throughout the literature. 
This procedure is applicable for any orientation distribution 
$\Theta$ that can be parameterized by a vector $\bm{\alpha}$.

Recall that $A^*$ is the sorted adjacency matrix of a random nanowire
network and has size $N+2 \times N+2$. 
The first $N$ indices correspond to a nanowire, sorted by increasing
$x$-coordinate, while indices $N+1$ and $N+2$ correspond to the left
and right border electrode. 
The elements of the expected adjacency matrix $\mathrm{E}A^*$ are
thus the probability of intersection between the objects of indices 
$i,j$. 
To compute the matrix, we thus need to compute the probability of 
intersection between every pair of wires, conditioned on the 
rank of the $x$-coordinate of each wire. 
We also need to calculate the probability of intersection between 
each wire and the border electrodes, conditioned on the rank of the 
$x$-coordinate of the wire. 
Because the matrix is symmetric, we only need to do so for $i > j$, 
and we only need to do the calculation for a single border 
electrode because the probabilities for the other border electrode 
are symmetric.

We will first calculate the probability of intersection between any 
two nanowires.
Denote a wire as $w=(x,y,\theta)$ and let $w^*_i,w^*_j$
be the $i$th and $j$th wire according to the sorted
order based on $x$-coordinate.
The desired probability is then denoted by $\mathrm{P}(w^*_i \cap w^*_j)$.
The event of $w^*_i$ intersecting $w^*_j$ is a deterministic
function of the difference in $x$ coordinates, the difference
in $y$ coordinates, and the angles of the two wires.
Under our independence assumptions, we can thus calculate
$\mathrm{E}A^*_{ij}$ by calculating the distributions of
$x^*_i - x^*_j$ and $y^*_i - y^*_j$ and then using the
known distributions of $\theta_i$ and $\theta_j$.
For brevity, we define
\begin{eqnarray}
    x_{ij} && = x_{i}^{*}-x_{j}^{*} \\
    y_{ij} && = y_{i}^{*}-y_{j}^{*}
\end{eqnarray}

We will first analyze randomness solely in $y_{ij}$ by 
computing the intersection probability conditioned on $x_{ij}$, 
denoted by $\mathrm{P}(w^*_i \cap w^*_j | x_{ij})$.
This is the probability that two wires intersect if
we know the difference in $x$-coordinates between them. 
For any pair of wires $w^*_i, w^*_j$ with $x$-separation $x_{ij}$ 
and angles $\theta_i,\theta_j$, we can define the horizontal range
of overlap $b$ as the length of the interval of $x$-coordinates 
that both wires lie in. 
For particular values of $b, \theta_i, \theta_j$, there is an
interval of $y_{ij}$ values for which $w_i$ and $w_j$ will cross.
We denote the length of this interval of by $h$.
We illustrate these quantities with example nanowire pair
configurations in Figure 3.

\begin{figure} 
\includegraphics[width=\linewidth]{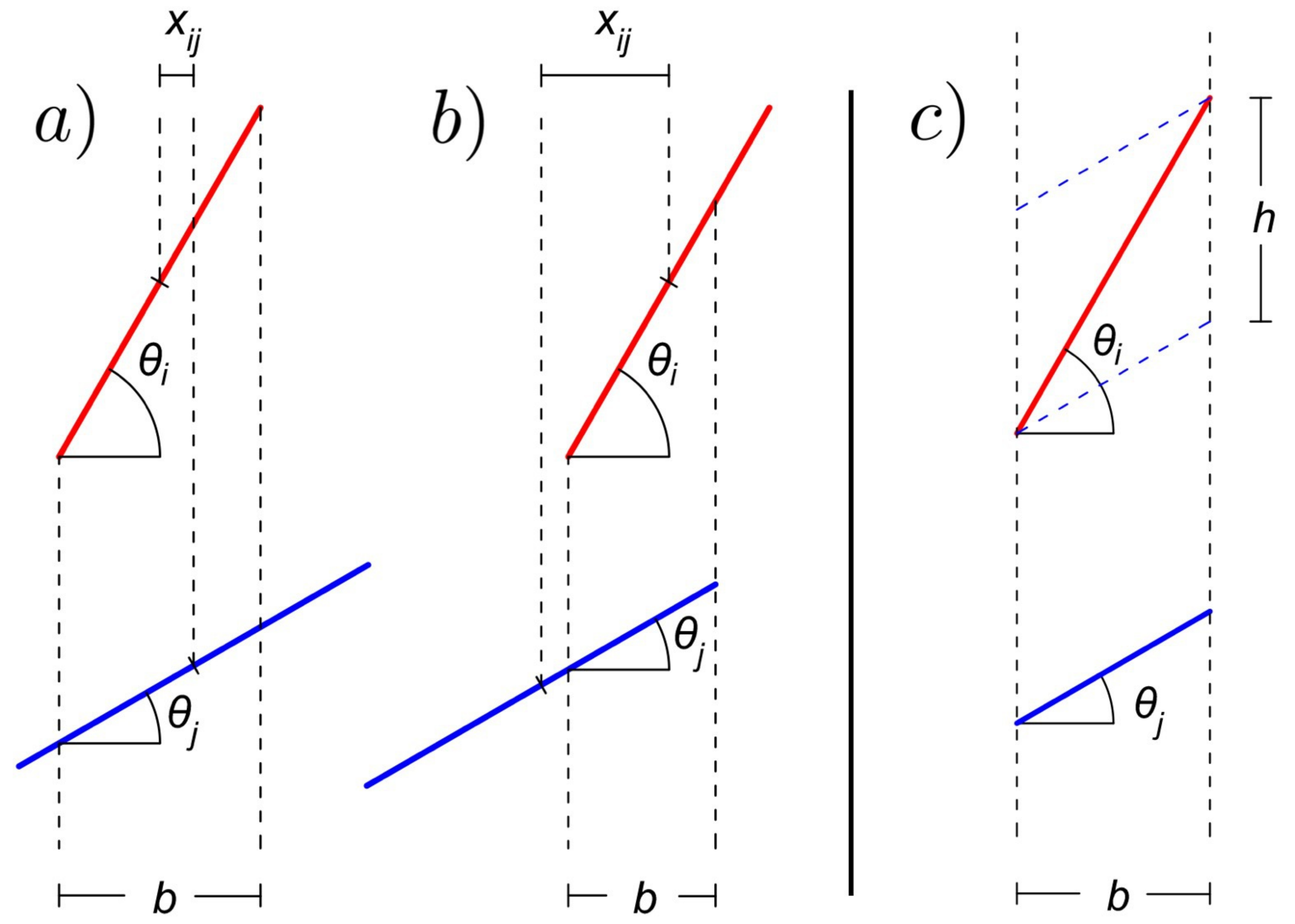}
\caption{If two nanowires have known $x$-separation of $x_{ij}$
and angles $\theta_i$ and $\theta_j$, we can calculate the 
length of the horizontal region of overlap $b$.
We visualize this quantity in the case where the horizontal
region of one nanowire is contained in the region of the other 
(a) as well as the case where this is not true (b).
We can then use this quantity to calculate the length of the
range of $y$-separations $h$ for which the two 
nanowires would intersect as visualized in (c). } 
\end{figure}

Now observe that the distribution of $y_{ij}$ is identical for all
$i \neq j$.
Furthermore, because of our use of periodic boundary conditions,
since $\mathcal{Y}$ is the uniform distribution, $y_{ij}$ is in fact
the uniform distribution in the range $[0, 1)$. Therefore, the
probability of intersection between $w_i$ and $w_j$ conditioned
on $x_{ij}$ is given by the conditional expectation of $h$:
\begin{align}
    \mathrm{P}(w^*_i \cap w^*_j | x_{ij}) &= \mathrm{E}[h | x_{ij}]
\end{align}
We can calculate this conditional expectation by observing
that $b$ and $h$ can be calculated from $x_{ij}$ and 
$\theta_i,\theta_j$, as in 
Eq.~(\ref{b_equation}, \ref{h_equation}). 
Here, $(f)_+$ is defined to be $\max\{f,0\}$.
\begin{align}
b\left(x_{ij},\theta_{i},\theta_{j}\right) & =\min\begin{cases}
((l/2)\left(\cos\theta_{i}+\cos\theta_{j}\right)-x_{ij})_+\\
l\cos\theta_{i}\\
l\cos\theta_{j}
\label{b_equation}
\end{cases}\\
h\left(x_{ij},\theta_{i},\theta_{j}\right) &%
=b\left(x_{ij},\theta_{i},\theta_{j}\right)%
\left|\tan\theta_{i}-\tan\theta_{j}\right|
\label{h_equation}
\end{align}
In Eq.(\ref{b_equation}), the latter two cases correspond to the
situation when the interval of $x$-coordinates that one wire lies in
is contained by the interval of $x$-coordinates that the other
lies in, as in Figure 3a.
The first case is taken when this situation does not occur, as in
Figure 3b.

The conditional expectation is then given by integrating out
$\theta_i$ and $\theta_j$ drawn independently from $\Theta$:
\begin{align}
    \mathrm{P}(w^*_i \cap w^*_j | x_{ij}) &= %
    \int h(x_{ij},\theta_{i},\theta_{j}) %
    p(\theta_i)p(\theta_j) d(\theta_i\theta_j)
    \label{cond_prob}
\end{align}
Since we have assumed the wires are sorted by $x$-coordinate,
the difference $x_{ij}$ is the difference in order
statistics $i$ and $j$ from the distribution $\mathcal{X}$.
Because $\mathcal{X}$ is the uniform distribution, 
$x_{ij}$ follows the Beta distribution with parameters
$i - j$ and $N - i + j + 1$, if $i > j$ 
\cite{weisberg1971distribution}.
However, for the networks with large $N$ which we study, these
distributions become strongly concentrated at their mean, which is
$\frac{i-j}{N+1}$.
We thus assume that $x_{ij}$ is equal to its expected value, and
empirically observe no loss in accuracy.
This yields a formula for the probability of intersection between
any two nanowires:
\begin{align}
    \mathrm{P}(w^*_i \cap w^*_j) &= %
    \mathrm{P}(w^*_i \cap w^*_j | x_{ij} = \frac{i -j}{N+1})
\end{align}

A similar argument can be used to calculate the probability that
any wire crosses the left border electrode, denoted by $e_1$. 
Observe that $w^*_i$ intersects $e_1$ if and 
only if $(l/2) \cos \theta_i \geq x_i^*$. Assuming that $x_i^*$ equals
its expected value of $\frac{i}{N+1}$, the desired probability 
is then given by Eq.~(\ref{wire_electrode}), where $\theta \sim \Theta$.
\begin{align}
    \mathrm{P}(x_i^* \cap e_1) &= %
    \mathrm{P}(\cos \theta \geq \frac{2i}{l(N+1)})
    \label{wire_electrode}
\end{align}
We can therefore calculate every element of $\mathrm{E}A^*$ for 
any $N$ and any orientation distribution $\Theta$, assuming 
wire positions are uniform. 

To use these expressions efficiently, we numerically compute the 
integral in Eq.~(\ref{cond_prob}) over a grid of values for 
$x_{ij}$ and the parameters $\bm{\alpha}$ of the orientation 
distribution $\Theta$.
We then fit a polynomial to the probability values on these grid points 
to obtain an expression  for $\mathrm{P}(w^*_i \cap w^*_j)$ that 
is extremely rapid to use. 
We further describe the speed of our method in the next subsection.

\subsection{Analysis of computational speed}
One of the significant advantages of our method is that it replaces
Monte Carlo sampling with an asymptotically faster computation.
Sampling-based models, which are the most common approaches for studying
random nanowire networks, have two major components.
First, a number of nanowire networks $M$ are sampled by directly
sampling $(x,y,\theta)$ for each of $N$ nanowires, and a collection
of $M$ adjacency matrices are calculated. 
Second, the Kirchhoff current equation is solved for each adjacency
matrix to collect $M$ observations of sheet conductivity, and these
observations are then averaged. 
The first of these steps has complexity $O(MN^2)$. 
Within all networks, each of the $N$ nanowires must be compared 
with a fixed fraction of all other nanowires for intersection to 
compute the adjacency matrix $A$.
The second step, meanwhile, has complexity $O(MN^3)$, which is the
cost of solving $M$ linear systems of equations each involving
$N$ variables. 
While the second step has larger complexity, both
steps require significant amounts of time and so
speeding up either is beneficial. 

Our model $\sigma^*$ delivers a large asymptotic improvement to 
the first step and delivers a large constant factor improvement to
the second step.
Recall that the probability of intersection between two wires under
our model depends only on the expected $x$-separation between them.
As a result, we only need to directly compute two rows of
$\mathrm{E}A^*$ in order to produce the entire matrix. 
This is because the expected $x$-separation between the wires
$w_i^*$ and $w_j^*$ is determined completely by the quantity
$|i - j|$. 
Equivalently, if we ignore the rows representing electrodes, then 
all diagonals of $\mathrm{E}A^*$ are constant.
We therefore must compute the first row of $\mathrm{E}A^*$ to 
obtain the probability of interaction between every pair of
nanowires, and also must compute the last row of $\mathrm{E}A^*$ to
obtain the probability of interaction between every nanowire and
an electrode. 
Therefore, the cost of computing $\mathrm{E}A^*$ is $O(N)$
with constant proportional to the time it takes to compute 
$\mathrm{P}(w^*_i \cap w^*_j)$.
We must still solve a single Kirchhoff current equation, and this
step is $O(N^3)$. 

Numerically integrating to compute each evaluation of
$\mathrm{P}(w_i^* \cap w_j^*)$ is in practice quite slow.
We therefore precompute this function for a grid of values of
$x_{ij}$ as well the parameters $\bm{\alpha}$ of the orientation
distribution $\Theta$, and then fit a polynomial to the computed
values.
A polynomial fit is in practice quite accurate because the probability
in question is smooth as a function of the parameters of interest.
This step makes computation of $\mathrm{P}(w_i^* \cap w_j^*)$
extremely rapid, but the precomputation cost is exponential in 
the number of parameters of the orientation distribution. 
For the majority of interesting cases, the orientation distribution
can be parameterized in one or two parameters, and this complexity 
is thus not significant compared to other steps.

In total, our method has a small precomputation cost, but replaces
the $O(MN^2)$ complexity of Monte Carlo sampling with an 
asymptotically faster $O(N)$ computation.
It also reduces the cost of solving linear systems by a factor of $M$,
the number of samples that are collected in a sampling based approach.
In our implementation, this allowed the model $\sigma^*$ to be 
evaluated about 100 times faster than direct Monte Carlo sampling. 

\begin{figure} 
\includegraphics[width=\linewidth]{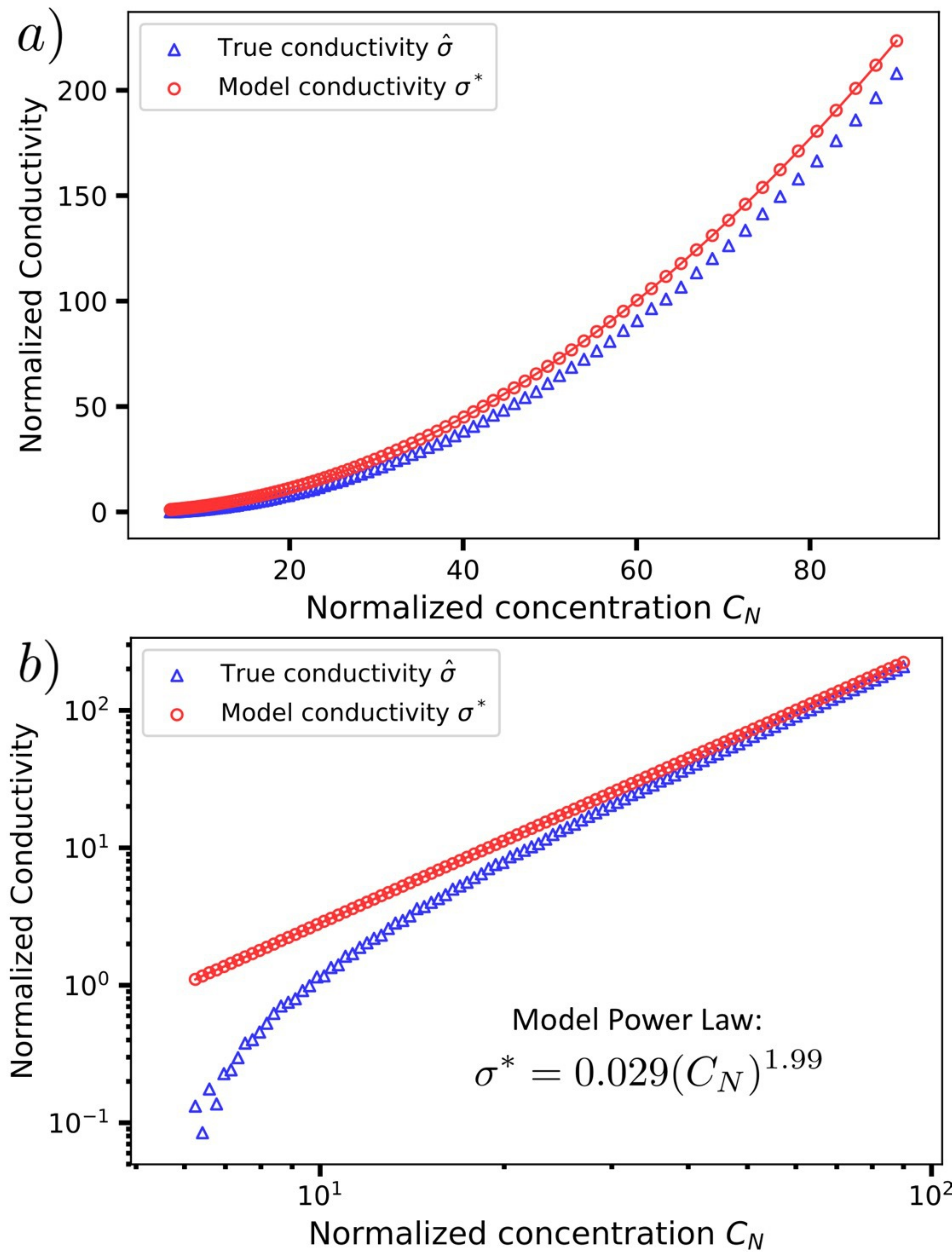}
\caption{Normalized conductivity as a function of normalized
wire concentration is shown for both the true empirical mean
and our approximate analytical model, on a linear scale (a) and a
double log scale (b). The model follows an
exact power law relationship and corresponds to the high density
behavior of the true conductivity. We emphasize that the
model is not obtained by fitting to the true conductivity.} 
\end{figure}

\section{Empirical Tests and Discussion}
We examine the effectiveness of $\sigma^*$ in modeling dependency
of network conductivity on both nanowire density and orientation
distribution. 
We assume, as in the previous section, that $\mathcal{X},
\mathcal{Y}$ are both the uniform distribution.
We implemented direct sampling of $\sigma$ under this assumption
and our previously stated setting, while allowing the
distribution $\Theta$ to be arbitrary.
We use our implementation of direct sampling of $\sigma$
as a baseline comparison for all tests, estimating
$\mathrm{E}\sigma$ with $\hat{\sigma}$ with $M=30$. 
Throughout these experiments, we use $l = 0.1$.
Larger values of $M$ reduce the noise of $\hat{\sigma}$, 
while smaller values of $l$ reduce finite size error in  
$\hat{\sigma}$ and $\sigma^*$.
The chosen values of $M$ and $l$ were found empirically to be 
sufficient to largely eliminate these errors;
$\sigma^*$ and $\hat{\sigma}$ do not change much for higher $M$
or lower $l$, as long as $C_N$ is fixed.

\subsection{Dependence on nanowire density}
We first assume that $\Theta$ is the isotropic distribution
(uniform in $[-90, 90]$ degrees) and explore the dependence
of $\sigma^*$ on normalized concentration $C_N$.
Figure 4 shows a comparison between normalized $\sigma^*$
and $\hat{\sigma}$ as a function of $C_N$, starting just above
the percolation threshold, on both linear and double log scales.
Conductivities are normalized by multiplying by junction
resistance $1\Omega$ so that they are unitless.
It is a known result that $\mathrm{E}\sigma$ can be approximated
as a power law function of the distance of normalized
concentration from the percolation threshold, with an
exponent of around 1.75 at medium densities which moves close
to 2 at high densities \cite{mutiso2013integrating}. 
Our estimate $\hat{\sigma}$ matches these known relationships;
the growth pattern of conductivity in log-space becomes linear
as the subtraction of the percolation threshold becomes negligible.
Our model $\sigma^*$, however, displays a perfect power law
dependence on $C_N$, with an exponent that matches the
asymptotic exponent of $\mathrm{E}\sigma$.
Near the percolation threshold, the error is large,
as we have assumed nanowire density above this threshold in 
developing our model. 
However, the error in log space approaches zero as concentration 
grows, and the model can thus be interpreted as the limiting 
behavior of $\mathrm{E}\sigma$ at high concentrations.

While $\sigma^*$ is less precise than other recent models for
predicting dependency of conductivity on concentration at small 
nanowire densities, the result that our approach yields the 
correct limiting behavior is theoretically interesting. 
By using $\mathrm{E}A^*$, our model directly encodes clustering
of nanowire only in the $x$-direction. 
However, this is sufficient information to capture asymptotic 
behavior, and, as we next show, capture the effect of varying
orientation distribution. 

\begin{figure} 
\includegraphics[width=0.65\linewidth]{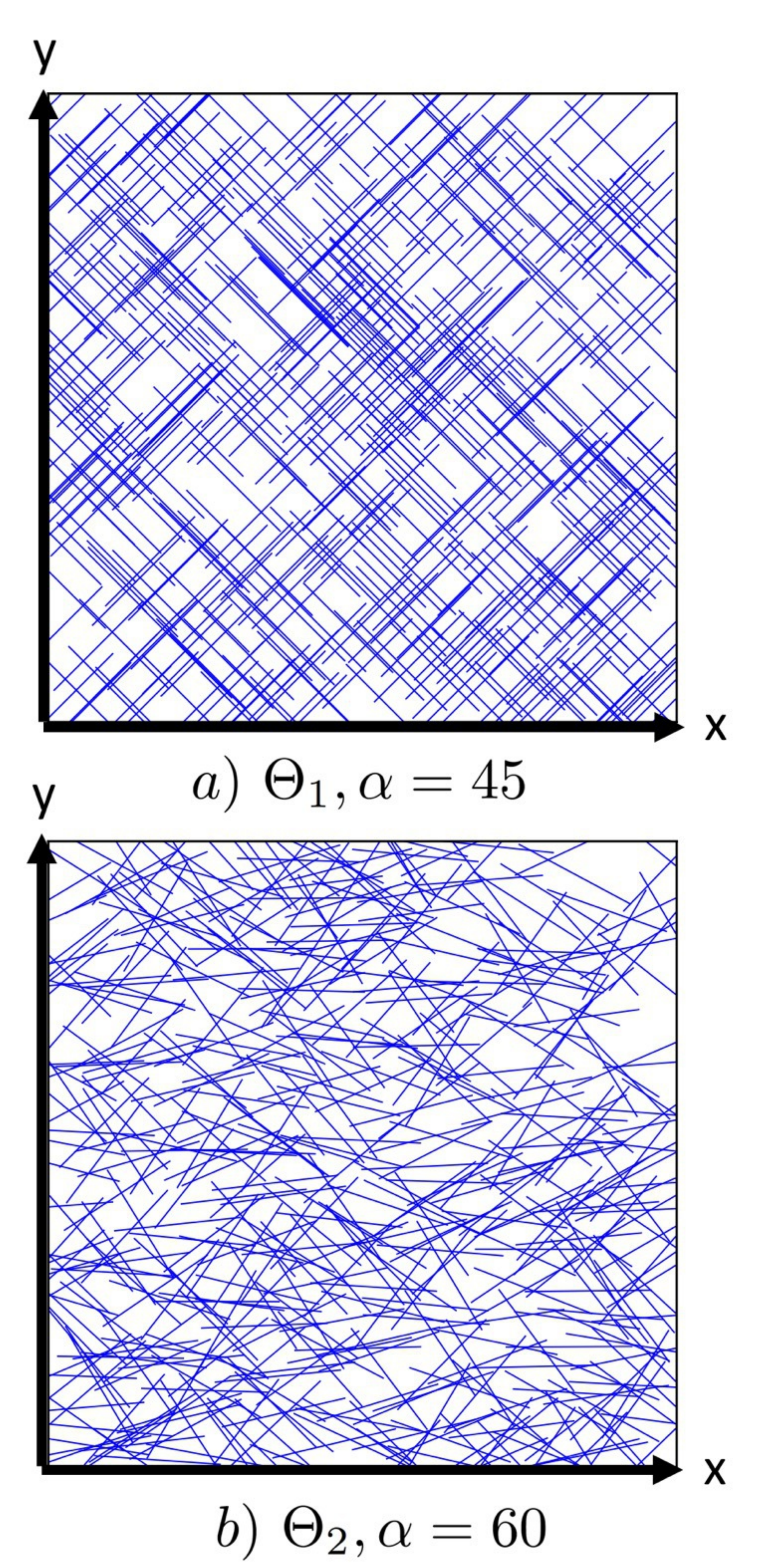}
\caption{ Nanowire networks drawn from (a) $\Theta_1$, with
$\alpha=45$ and (b) $\Theta_2$, with $\alpha=60$ are shown.
In $\Theta_1$, all nanowires have orientation at $\pm \alpha$
degrees from the horizontal ($x$-direction). In $\Theta_2$,
nanowire orientation is distributed uniformly in
$[-\alpha,\alpha]$ degrees from the horizontal.} 
\end{figure}

\subsection{Dependence on orientation distribution}

Our model $\sigma^*$ is particularly valuable because it is able 
to predict the effect of arbitrary orientation distributions on 
sheet conductivity. 
The problem of optimizing orientation distribution in random 
nanowire networks has been studied numerous times via computational
models, but there is no unified understanding of the observed 
effects
\cite{behnam2007effects, du2005effect, white2009simulations,%
      pimparkar2007limits, jagota2015conductivity}.

We consider two families of distributions for $\Theta$, each of
which is described by a single parameter. 
For each family, we demonstrate that $\sigma^*$ accurately captures
the effect of varying the distribution parameter on conductivity.
The first family $\Theta_1(\alpha)$ is given by
\begin{equation}
\label{fam_delta}
p(\theta) =
\begin{cases}
\frac{1}{2} & \text{$\theta = \alpha$} \\
\frac{1}{2} & \text{$\theta = -\alpha$}
\end{cases}
\end{equation}
for all $0 < \alpha < 90$.
All probability mass is concentrated at $\pm \alpha$
degrees from horizontal.
The second family $\Theta_2(\alpha)$ is given by 
\begin{equation}
\label{fam_unif}
p(\theta) =
\begin{cases}
\frac{1}{2\alpha} & \text{$|\theta| < \alpha$} \\
0 & \text{o.w.}
\end{cases}
\end{equation}
Probability density is uniformly distributed over
$[-\alpha, \alpha]$ degrees.
Figure 5 shows a sample network from a single distribution
within each family.
These two families were previously studied, and it was
found that while a conductivity gain over isotropic networks
could be achieved within $\Theta_2$, no gain could
be achieved within $\Theta_1$ \cite{jagota2015conductivity}.

\begin{figure} 
\includegraphics[width=\linewidth]{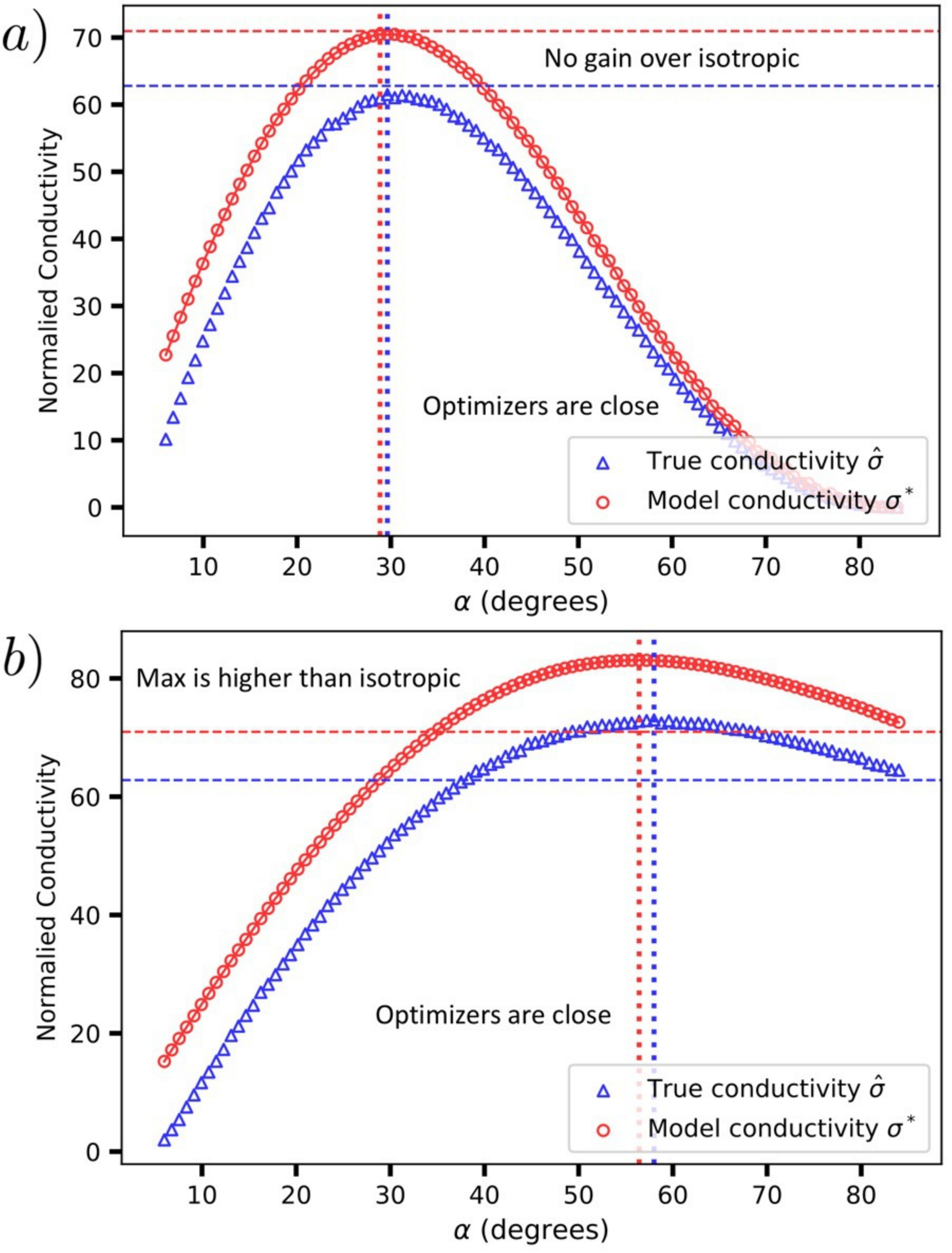}
\caption{ Normalized conductivity is shown as a function of
distribution parameter $\alpha$ for both the true empirical
conductivity and the approximate analytical model, for both
(a) $\Theta_1$ and (b) $\Theta_2$, with $C_N = 50$. In both 
families, the shapes of the two curves match well and the 
optimal values (vertical lines) are close. The model also
captures the fact that a gain over isotropic conductivity 
(horizontal lines) can only be achieved in $\Theta_2$.} 
\end{figure}

Figure 6 shows a comparison between $\sigma^*$ and
$\hat{\sigma}$ for determining the relationship between
distribution parameter $\alpha$ and normalized conductivity
for both $\Theta_1$ and $\Theta_2$.
The normalized concentration is fixed at 50 in both cases.
Within both families, the shape of the curve matches well, 
and the optimal values are within a few degrees of each other.
Moreover, the predictions from $\sigma^*$ match
$\mathrm{E}\sigma$ in that a gain over isotropic orientation
is attainable in $\Theta_1$ but not $\Theta_2$.

To the best of our knowledge, our model is the first to accurately
reproduce the effects of orientation distribution on sheet
conductivity without relying on Monte Carlo sampling in any
capacity.
These results indicate that orientation effects can be modeled
by analyzing their effects on network connectivity in a single
direction, as our model $A^*$ only takes into account positions
of nanowires in the $x$-direction.

\section{Conclusion}
We developed an approximate analytical model for sheet
conductivity of random nanowire networks that condenses a
large amount of their structure through a
specific choice of nanowire permutation. 
We showed that this model is an upper bound and matches
the asymptotic dependency of the true sheet conductivity
on wire concentration. 
We also demonstrated that the model accurately captures
the effects of orientation on nanowire network conductivity,
a result that has limited theoretical explanation in the literature. 
Our model is the first to accurately capture the effects of 
arbitrary orientation distributions on network conductivity,
and replaces Monte Carlo sampling with an asymptotically faster
computation.
These results and the structure of the model we
developed provide novel theoretical intuition about random nanowire
network conductivity.
Namely, our results demonstrate that network connectivity in the 
direction of current flow is the key factor in determining the 
dependence of conductivity on wire density and orientation 
distribution, because our model only encodes connectivity 
information in the $x$-direction.

The most pressing direction for future research is to
relax our assumption of zero wire resistance, as recent work has
indicated that the junction resistance in silver nanowire
networks can be reduced to a comparable magnitude as the
wire resistance \cite{bellew2015resistance}.
This could be done, for example, by using an approximate function
to calculate sheet conductivity based on $\mathrm{E}A^*$ in the
presence of wire resistance. 
Various recent analytical models for random nanowire network conductivity
have successfully used approximations about the number of
nanowires that a given nanowire will intersect
\cite{kumar2017current, forro2018predictive}.
Rather than using approximations derived in the setting of uniform
wire orientation, these models could instead use approximations
obtained from $\mathrm{E}A^*$ for an arbitrary orientation distribution.
The success of these existing models indicates that they would likely
function as accurate approximate functions to calculate sheet
conductivity given the information in $\mathrm{E}A^*$.

%\bibliographystyle{apsrev4-2}
%\bibliography{nanowires}% Produces the bibliography via BibTeX.
\input{nanowires.bbl}
\end{document}

%% file: nanowires.bbl
\providecommand{\noopsort}[1]{}\providecommand{\singleletter}[1]{#1}%